\documentclass[letterpaper]{article} 
\usepackage{aaai23}  
\usepackage{times}  
\usepackage{helvet}  
\usepackage{courier}  
\usepackage[hyphens]{url}  
\usepackage{graphicx} 
\usepackage{natbib}  
\usepackage{caption} 
\usepackage{algorithm}
\usepackage{algorithmic}
\usepackage{amssymb}
\usepackage{bbding}
\usepackage{amsmath}
\usepackage{multirow}
\usepackage{newfloat}
\usepackage{listings}
\DeclareCaptionStyle{ruled}{labelfont=normalfont,labelsep=colon,strut=off} 
\floatstyle{ruled}
\newfloat{listing}{tb}{lst}{}
\floatname{listing}{Listing}
\title{TESSP: Text-Enhanced Self-Supervised Speech Pre-training }
\author {
    Zhuoyuan Yao, \textsuperscript{\rm 1, \rm 2}
    Shuo Ren, \textsuperscript{\rm 2}
    Sanyuan Chen, \textsuperscript{\rm 2}
    Ziyang Ma, \textsuperscript{\rm 2}
    Pengcheng Guo, \textsuperscript{\rm 1}
    Lei Xie
}
\affiliations {
    \textsuperscript{\rm 1}Audio, Speech and Language Processing Group (ASLP@NPU), School of Computer Science\\
    Northwestern Polytechnical University, Xi'an, China \\
    \textsuperscript{\rm 2}Microsoft Corporation \\
    zhyyao@npu-aslp.org
}

\begin{document}

\maketitle

\begin{abstract}

Self-supervised speech pre-training empowers the model with the contextual structure inherent in the speech signal while self-supervised text pre-training empowers the model with linguistic information. Both of them are beneficial for downstream speech tasks such as ASR. However, the distinct pre-training objectives make it challenging to jointly optimize the speech and text representation in the same model. To solve this problem, we propose \textbf{T}ext-\textbf{E}nhanced \textbf{S}elf-Supervised \textbf{S}peech \textbf{P}re-training (TESSP), aiming to incorporate the linguistic information into speech pre-training. 
Our model consists of three parts, i.e., a speech encoder, a text encoder and a shared encoder. The model takes unsupervised speech and text data as the input and leverages the common HuBERT and MLM losses respectively. 
We also propose \textbf{phoneme up-sampling} and \textbf{representation swapping} to enable joint modeling of the speech and text information.
Specifically, to fix the length mismatching problem between speech and text data, we phonemize the text sequence and up-sample the phonemes with the alignment information extracted from a small set of supervised data. Moreover, to close the gap between the learned speech and text representations, we swap the text representation with the speech representation extracted by the respective private encoders according to the alignment information. 
Experiments on the Librispeech dataset shows the proposed TESSP model achieves more than 10\% improvement compared with WavLM on the test-clean and test-other sets. We also evaluate our model on the SUPERB benchmark, showing our model has better performance on Phoneme Recognition, Acoustic Speech Recognition and Speech Translation compared with WavLM.

\end{abstract}

\section{Introduction}
In the past years, self-supervised learning has achieved great success in natural language processing (NLP). The typical self-supervised learning frameworks, such as BERT \cite{devlin2018bert} and BART \cite{lewis2020bart}, propose the Masked Language Modeling (MLM)~\cite{taylor1953cloze} pretraining method to learn the universal text representation from large-scale unlabeled text data, which brings significant performance improvements on almost all the NLP tasks.

Inspired by the success of self-supervised learning in NLP, a series of self-supervised learning methods for speech processing are introduced and show promising results on a wide range of speech tasks \cite{baevski2020wav2vec,hsu2021hubert,chen2022wavlm}. For instance, HuBERT learns the universal speech representation with a BERT-like masked prediction method. The model is fed with the masked speech feature and trained to predict the discrete pseudo labels of the masked regions, where the discrete pseudo labels are generated by an offline clustering step.

Despite the great success, those self-supervised learning methods only focus on either text information modeling or speech information modeling during pre-training.
Recently, the self-supervised learning methods for joint speech and text modeling have attracted increasing attention \cite{kim2021st,qian2021speech,ao2021speecht5,zheng2021wav,bapna2021slam,chen2022maestro}. However, current methods utilize either the encoder-decoder model or the transducer architecture, and show limited performance improvement in the down-streaming speech tasks such as acoustic speech recognition (ASR).
In this work, we propose \textbf{T}ext-\textbf{E}nhanced \textbf{S}elf-Supervised \textbf{S}peech \textbf{P}re-training (TESSP), a novel Transformer encoder-based joint speech and text pre-training framework, to enhance the self-supervised speech modeling by leveraging large amounts of unlabeled text data and a small amount of supervised speech-to-text data. Our TESSP consists of a speech encoder, a text encoder, and a shared encoder. The speech and text encoders observe speech and text input respectively, and their outputs are fed into the shared encoder.
According to the type of the input data, we divide training tasks into three types as follows:
\begin{enumerate}
    \item With the speech encoder and shared encoder, the speech pre-training task aims to map the speech input to the semantic space and the model is trained with the loss similar to HuBERT.
    \item With the text encoder and shared encoder, the text pre-training task aims to convert the phoneme sequence input to the text output.  A phoneme-level MLM loss and a character-level Connectionist Temporal Classification (CTC) loss are introduced to train the model. To fix the length mismatching problem between speech and text data, we propose \textbf{phoneme up-sampling} to phonemize the text sequence and up-sample the phonemes with the alignment information extracted from a small set of supervised data
    \item With all the three encoders, the paired data pre-training task forces the model to map the different modalities into a shared semantic space, leveraging a small set of paired speech-text data. To learned the alignment between speech and text representations implicitly, we propose \textbf{representation swapping} to swap the text representation with the speech representation extracted by the respective private encoders according to the alignment information, and take the mixed representation as the input into the shared encoder.
\end{enumerate}

From experiments, we find that TESSP learns the alignment between speech and text data and the shared representation space for the two modalities in middle layer, which makes TESSP perform better in ASR tasks. We evaluate our model on the LibriSpeech~\cite{panayotov2015librispeech} ASR task, which shows that our framework can achieve over 10\% relative word error rate (WER) reduction compared to the strong baselines even with language model fusion. Our further evaluation on SUPERB \cite{yang2021superb} also shows that TESSP has better performance on Phoneme Recognition (PR), Acoustic Speech Recognition (ASR), Speech Translation (ST) compared with WavLM~\cite{chen2022wavlm}.

\section{Related Work}
\subsection{Text pre-training}
Self-supervised learning of language representations using neural networks has a long history. Word2vec~\cite{mikolov2013distributed} started to train word representations from unannotated data by noise contrastive estimation techniques~\cite{gutmann2012noise, mnih2012fast}. A series papers follow Word2vec focus on expanding the approach to contextual representations of sentences, including ELMo~\cite{sarzynska2021detecting}, GPT~\cite{radford2018improving}, BERT~\cite{devlin2018bert} and T5~\cite{raffel2020exploring}. These methods rely on either generative language modelling~\cite{bengio2003neural} or MLM. The self-supervised pre-training approaches have achieve significant improvements on a wide variety of downstream tasks~\cite{wang2019superglue, hu2020xtreme}
\subsection{Speech pre-training}
Similar to text pre-training, the approaches were explored in speech pre-training. The word2vec approach is used in~\cite{chung2016audio} to learn vector representations of variable-length audio segments. 
\cite{oord2018representation} introduces contrastive predictive coding (CPC) which leverages language modeling an negative sampling to learn speech representation.
CPC was followed by a series of papers called wav2vec model. 
The first wav2vec model~\cite{schneider2019wav2vec} closely follow the CPC approach using a contrastive binary classification task for unsupervised pre-training. vq-wav2vec~\cite{baevski2019vq} using Gumbel softmax~\cite{jang2017categorical} to simulate a vector quantizer inspired by VQ-VAE~\cite{van2017neural}.
wav2vec 2.0~\cite{baevski2020wav2vec} merges the quantization and contrastive learning into a unified learning procedure that pre-trains a Transformer model. wav2vec 2.0 shows significant gains on Librspeech.
The quantized speech representation change the speech utterances into sequences of discrete tokens belonging to a fix vocabulary, similar to text, where MLM approach can be applied.
HuBERT~\cite{hsu2021hubert} uses an offline k-means clustering to quantize the speech representations to target labels for a BERT-like prediction loss.
Based on HuBERT, WavLM~\cite{chen2022wavlm} jointly learns masked speech prediction and denoiseing by adding noise to the input speech data. WavLM trains a Transformer model with gated relative position bias.
With further using of the mask method, data2vec~\cite{baevski2022data2vec} predict latent representations of the full input based on a masked view of the input in a self-distillation architecture with Transformer model. Data2vec achieve significant improvement on Librispeech.

\subsection{Joint speech and text pre-training}
There are many works try to jointly training unpaired speech and text data recently. Previous work \cite{kim2021st} and ~\cite{qian2021speech} improve the model performance on speech Language understand(SLU) task. 
SpeechT5~\cite{ao2021speecht5} leverage an encoder-decoder architecture to pre-train both unsupervised speech data and text data in one model.  
Wav2vec-U~\cite{baevski2021unsupervised} and wav2vec-U 2.0 ~\cite{liu2022towards} segment the audio representations and introduce a GAN network to learn the representations of audio and phoneme sequences into the same space to optimizing pre-trained model performance.
Wav-BERT~\cite{zheng2021wav} and \cite{wang2022optimizing} fuse BERT into an ASR model. Wav-BERT introduces a representation aggregation module to generate new representations for speech and text encoder outputs, while ~\citet{wang2022optimizing} shares some encoder layer parameters for speech and text and adopts a swap strategy. These works show that learning common representations of speech and text makes sense in speech recognition which motivates us to fuse the text data in speech self-supervised pre-training.
SLAM~\cite{bapna2021slam} and mSLAM~\cite{bapna2022mslam} train the shared encoder with self-supervised training loss on unlabeled speech and text data and propose alignment loss on paired speech and text data.
MAESTRO~\cite{chen2022maestro} incorporates text into speech pre-training and uses paired data to assist speech and text joint training. However, during paired speech and text training, MAESTRO introduce the down-stream speech recognition learning target which learn the speech and text data to a shared representation space explicitly. In our work, we will implicitly guide the model to learn a common representation space for text and speech.

\section{Methodology}
\subsection{Overview\label{model architecture}}
\begin{figure}[t]
\centering
\includegraphics[width=1.0\columnwidth]{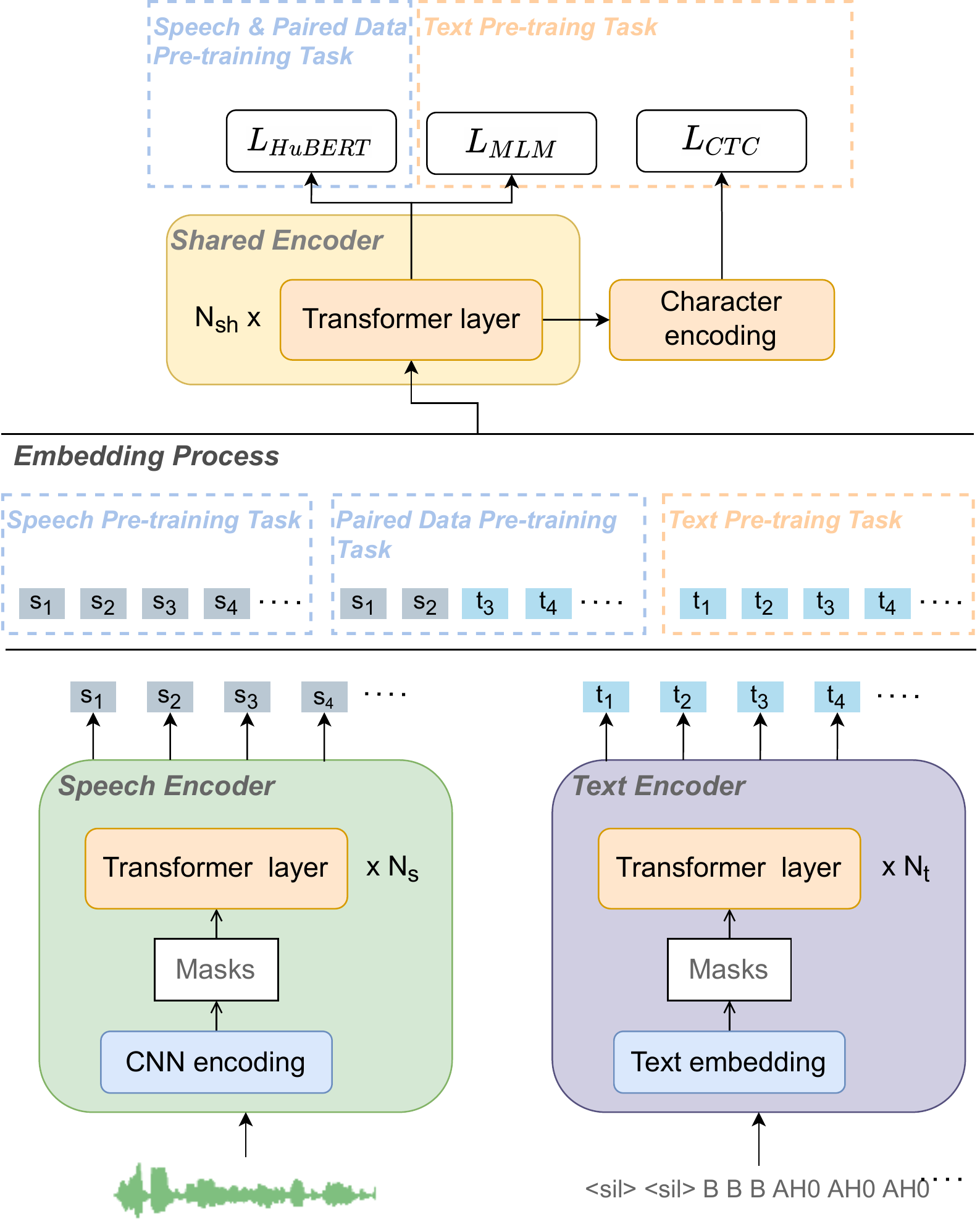} 
\caption{The architecture of TESSP consists of three parts, a speech encoder, a text encoder, and a shared encoder. An embedding process module is introduced between the private encoders and the shared encoder to process the encoder outputs. We design three pre-training tasks according to different types of input data, i.e., the speech pre-training task, the text pre-training task, and the paired data pre-training task. 
}
\label{fig1}

\end{figure}

The proposed framework for joint speech and text pre-training is presented in Figure~\ref{fig1}. It mainly consists of three parts: a speech encoder, a text encoder, and a shared encoder. The speech encoder extracts speech embeddings from the input speech signals while the text encoder generates phoneme embeddings from the input up-sampled phoneme sequences. The embeddings of the two modalities are processed according to different pre-training tasks, and then they are passed through the shared encoder to calculate the corresponding loss. We propose three pre-training tasks according to different input data. 

For the speech pre-training task, the model takes the masked unlabeled speech data as the input. The output embedding of the speech encoder is directly fed into the shared encoder. The final output of the shared encoder is used to calculate the HuBERT loss.

For the text pre-training task, we take the masked phonemized text data as the input. Note that the speech input sequence is always much longer than the phonemized text input sequence. Therefore, to encourage the shared encoder to map the speech and text into a shared space and to avoid simply distinguishing the speech and text data by the sequence length, we propose the \textbf{phoneme up-sampling} technique to up-sample each phoneme according to its duration length observed from a small set of paired speech-text data. So that, the up-sampled phoneme sequence can be treated as a simulated speech frame sequence. Then we randomly mask the up-sampled phoneme sequence and take it as the input to the text encoder, with its output feeding into the shared encoder without extra processing. The output of the shared encoder is used to predict the masked phonemes in the input sequence for calculating the phoneme-level Masked Language Model loss ($L_{\text{MLM}}$). Besides, we add an extra character encoding layer on top of the shared encoder and take its output to predict the corresponding original character sequence for calculating the character-level CTC loss ($L_{\text{CTC}}$)~ \cite{graves2006connectionist}. 


For the paired data pre-training task, we first input the paired speech and up-sampled phonemized text data into the corresponding encoders to get their latent representations. Then, we propose \textbf{representation swapping} to encourage the model to map the representation into a shared space. Before the shared encoder, we first randomly choose some phonemes representations and swap them with the speech representations at the same positions. Then, the processed speech representations are fed into the shared encoder to calculate $L_{\text{HuBERT}}$. 


The following sections will describe the details of each pre-training task.


\subsection{Speech Pre-Training}
HuBERT ~\cite{hsu2021hubert} uses K-means to cluster speech features and treat the cluster ids as pseudo training labels. Following WavLM ~\cite{chen2022wavlm}, we also employ gated relative position bias ~\cite{chi2022xlm} in our model, which is encoded based on the offset between the ``key" and ``query" in the Transformer self-attention mechanism. During pre-training, the speech encoder consumes speech signal and outputs the speech representation $S=\{s_1,s_2,s_3,s_4,\dots\}$. $S$ is then fed into the shared encoder directly and generate finally shared representation of speech $H^s=\{h_1^s,h_2^s,h_3^s,h_4^s,\dots\}$.

The network is optimized to predict the discrete target sequence $Z = \{z_1, z_2, z_3, z_4, \dots \}$, where each $z_t \in [C]$ is a $C$-class categorical variable. The probability over codewords can be parameterized as
\begin{equation}
    p(c|h_t^s)=\frac{\exp(sim(Wh_t^s,e_c)/\tau)}{ {\textstyle \sum_{c'=1}^{C}}\exp(sim(Wh_t^s,e_{c'})/\tau ) } ,
\end{equation}
where $W$ is a projection matrix, $h_t^s$ is the high-level speech representation for step $t$, $e_c$ is the embedding for codeword $c$, $sim(a,b)$ computes the cosine similarity and $\tau = 0.1$ scales the logit. A key ingredient of HuBERT is that the prediction loss is only applied over the masked regions, forcing the model to learn a combined acoustic and language model over the continuous inputs.

\subsection{Text Pre-Training}
An obvious problem of speech and text joint training is length mismatch. To learn the shared representation space of speech and text, we should avoid the model distinguishing two modalities of data simply through length mismatch. To solve this problem, we first use a python based English grapheme-to-phoneme conversion toolkit to transform the text data into phoneme sequence. Then, a specific SIL tag is randomly inserted into the phoneme sequence to represent the silence frame. Next, we obtain the duration distribution of each phoneme from a small amount of paired data. Finally, according to the duration distribution, we up-sample each phoneme and get the up-sampled phoneme sequence. 

To leverage the text data, The up-sampled text tokens is processed with a text embedding and a 6-layer text Transformer encoder to generate text representations $T=\{t_1,t_2,t_3,t_4,\dots\}$. Similar to speech pre-training methods, $T$ is directly fed into the shared encoder to get the finally shared representations of text $H^t=\{h_1^t,h_2^t,h_3^t,h_4^t,\dots\}$. 

Then, MLM loss and CTC Loss are calculated in text pre-training. 
Firstly, Though a phoneme-level linear embedding, $H^t$ is used to predict the masked phone and calculate the cross-entropy loss $L_{MLM}$ at the mask position.
Secondly, $H^t$ are fed into a 1-layer character encoding layer and a character linear embedding to predict the original character sequence. The prediction of character is trained by CTC loss $L_{CTC}$. Finally, the final training objective is defined as the sum of two losses.

\subsection{Paired Data Pre-Training}
\begin{figure}[t]
\centering
\includegraphics[width=1.0\columnwidth]{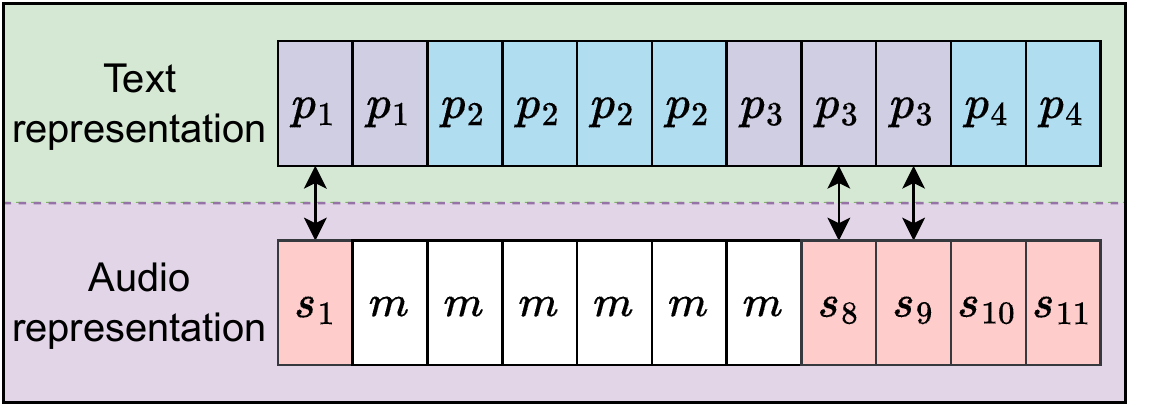} 
\caption{During paired data pre-training task, TESSP swap the speech representations with text representations. ${\{s_1,s_2,...,s_n\}}$ is speech representations sequence with mask embedding ${m}$ while ${\{p_1,p_2,...,p_n\}}$ is phoneme representations sequence. The swap strategy is adopted within phoneme boundary (the phoneme representations in purple) and avoid the mask embedding. }
\label{fig2}

\end{figure}

We find that it is hard for the model to learn the shared representation between speech and text with only unpaired speech and text data. So we leverage paired data to learn the alignment between the speech frames and the text input token. The alignment is done to get the phoneme label for each speech frame. Since the up-sample phoneme sequence is fed into text encoder, it is very intuitive to feed the aligned paired phoneme text data to the text encoder and get the text embedded features. Before shared encoder, as shown in Figure~\ref{fig2}, we randomly replace speech embedded features with their corresponding text embedded features on the unmasked speech time step according to the phoneme-aware strategy in \cite{wang2022optimizing}.

To verify whether the model can fuse the two modal data without modeling, we draw no supervised target loss during paired data pre-training. So the swapped representations features are directly used to predict the HuBERT target labels through a shared Transformer encoder.

\section{Experiments}
\subsection{Data}
For unsupervised speech pre-training, we use the full 960 hours of LibriSpeech \cite{panayotov2015librispeech} audio which is derived from the LibriVox project contains English recordings of copyright-free audiobooks by volunteers from the Internet. For unsupervised text pre-training, we use the text corpus of Librispeech, which contains 14500 public domain books. For paired data pre-training and supervised fine-tuning, we use LibriSpeech 100-hour(train-clean-100) split from the LibriSpeech training set. 
\subsubsection{Text processing}
Because of the length mismatch between the speech and the text data, the model can easily distinguish the data modal only according to the data length. However, it is expected to learn the shared representation space for the two data modalities. So we need to up-sample the text sequence. Firstly, we calculate the probability distribution of the repetition length of each phoneme from the paired aligned phoneme sequence. We find that there is a long tail problem in the statistical probability distribution. To solve this problem, we accumulate the probability from zero-times repetition and remove the part where the accumulated probability exceeds 0.98. Finally, we obtain the pairwise statistics of the repetition length and probability. Observing the paired alignment data, we find that there are SIL tags with different lengths in the alignment sequence. Therefore, inspired by ~\cite{baevski2021unsupervised}, we add the SIL tag to the beginning and the end of all phonemized unlabeled text sentences. Then we randomly insert SIL between groups of phonemes corresponding to words at a rate of 0.25. When generating text data, we sync the phoneme sequence and randomly sample the repetition length in the probability distribution to up-sample the phonemes and SIL tags encountered. Cause the longest repetition time of the phonemes is no more than 20, we choose the mask length of 40 to cover at least one whole phoneme token to avoid information leakage.
\subsubsection{Speech Label Generation}
The label generation follows~\cite{hsu2021hubert}. For the first iteration, we run k-means with 100 classes on 39-dimensional MFCC features over 960 hours training set and training by a normal Transformer encoder which hyperparameters follow ~\cite{hsu2021hubert}. Then we run k-means with 500 classes on the 6-th layer output features of the first iteration model.
\subsection{Pre-Training}
We adopt 6 layers of speech encoder, 6 layers of text encoder, 6 layers of the shared encoder and 1 layer of character encoding layer. Each layer is a Transformer block with model dimension 768, inner dimension 3072 and 8 attention heads. It is equipped with a convolution-based relative position embedding layer with kernel size 128 and 16 groups at the bottom. We also add a bucket relative position embedding which is mentioned in ~\cite{chen2022wavlm}.

For the speech pre-training task, we use the label generated from the first iteration model directly training the speech encoder and shared encoder. For masking, we randomly sample the starting positions with a probability of 0.08 and mask the subsequent 10 time steps. The masked spans may overlap.

For the text pre-training task, we randomly sample the mask starting positions with a probability of 0.15 and mask the subsequent 40 phonemes. The mask spans also may overlap. For Mask Language Model loss, we only compute the cross entropy at the masked position. For CTC loss, we use the character sequence corresponding to the text as the target label. To get a better result, the model needs to learn the phone representation first, so CTC loss is calculated after 50000 steps.

For paired data pre-training task, we use aligned text data as the input to the text encoder. Unlike text pre-training, we do not mask the phoneme sequence. For the high-dimensional representation of speech and text, we will accumulate the boundaries of each phoneme and swap on spans of unmasked frames that correspond to the same phoneme token in forced alignment.

Since the volume of our text data is much larger than that of unsupervised speech data, to avoid too much model bias towards text data, we will randomly sample the same batches of the unsupervised text data in each epoch of training. The complete set of unsupervised paired data is iterated in each epoch. Therefore, the model is trained for 835k steps. For both iterations, the model is trained on 16 V100 Tesla GPUs with a batch size of 87.5 seconds of audio for each GPU. We use the Adam optimizer and the learning rate ramps up linearly from 0 to 5e-4 for 32k steps in the first iteration but 64k steps in the second iteration. Then the learning rate decays linearly back to 0.

\subsection{Supervised Fine-Tuning and Decoding}
The model is fine-tuned on 8 GPUs with a batch size of 200 seconds of audio for each GPU. We optimize with Adam and a tri-stage rate schedule where the learning rate is warmed up for the first 10\% of updates, held constant for the next 40\% and then linearly decayed for the remainder. For evaluation, we use wav2letter++~\cite{pratap2019wav2letter++} beam search decoder with beam size 1500 for 4-gram language model (LM) fused decoding as: 
\begin{equation}
    \log p_{\text{CTC}}(y|x)+w_1logp_{\text{LM}}(y)+w_2|y|
\end{equation}
where $|y|$ means the length penalty of decoded output sequence.

\begin{table}[t]
\begin{center}
\begin{small}

\begin{tabular}{lccc}
\hline

\multirow{2}{*}{Model} & \multirow{2}{*}{LM} &   \multicolumn{2}{c}{WER$\%$($\downarrow$)}\\
& & test-clean & test-other \\ \hline

wav2vec 2.0 & None & 6.1 & 13.3 \\ 
HuBERT & None & 6.3 & 13.2 \\
WavLM & None &  5.7 & 12.0 \\
DeCoAR 2.0 & 4-gram & 5.0 & 12.1  \\
DiscreteBERT & 4-gram &  4.5 & 12.1 \\
wav2vec 2.0 & 4-gram  & 3.4 & 8.0 \\
HuBERT & 4-gram  & 3.4 & 8.1\\ 
WavLM  & 4-gram &  3.4 & 7.7\\ \hline
TESSP  & None &  5.10 & 10.88 \\ 
TESSP &  4-gram & 3.05 & 7.15   \\ 
\hline
\end{tabular}
\caption{Model comparisons in the \textsc{Base} setting. The performance of HuBERT w/o LM is obtained by fintuning the public released model. }
\label{main result}
\end{small}
\end{center}

\end{table}

\begin{table*}[h]
\centering

\resizebox{1.0\textwidth}{!}{
\begin{tabular}{l|c|c||c|c|c||c|c|c|c||c|c|cc||c}
\hline
\multirow{3}{*}{Method} & \multirow{3}{*}{\#Params} & \multirow{3}{*}{Corpus}
 & \multicolumn{3}{c||}{Speaker} & \multicolumn{4}{c||}{Content} & \multicolumn{4}{c||}{Semantics}  & ParaL \\ \cline{4-15}

& & & SID & ASV & SD & PR & ASR & KS & QbE & ST & IC & \multicolumn{2}{c||}{SF} & ER  \\ \cline{4-15} 

& & & Acc $\uparrow$ & EER $\downarrow$ & DER $\downarrow$  & PER $\downarrow$ & WER $\downarrow$ & Acc $\uparrow$ & MTWV $\uparrow$ & BLEU $\uparrow$ & Acc $\uparrow$ & F1 $\uparrow$ & CER $\downarrow$ & Acc $\uparrow$ \\ \hline 
wav2vec 2.0 Base & 95.04M & LS 960 hr & 75.18  & 6.02 & 6.08 & 5.74 & 6.43 & 96.23 & 0.0233 & 14.81 & 92.35 & 88.30 & 24.77 & 63.43 	 \\
HuBERT \textsc{Base} & 94.68M & LS 960 hr &  81.42 &  5.11 &  5.88  &      5.41        &    6.42       & 96.30 & 0.0736 &       15.53     &  98.34 & 88.53 &      25.20      & 64.92 \\
WavLM Base  & 94.70M & LS 960 hr & \textbf{84.51} & \textbf{4.69} & \textbf{4.55} & 4.84 & 6.21 & \textbf{96.79} & \textbf{0.0870} & 20.74 & \textbf{98.63} & \textbf{89.38} & \textbf{22.86} & \textbf{65.94} \\ \hline
TESSP  & 94.70M & LS 960 hr  &    77.37        &     6.15       &   6.07         & \textbf{4.58} & \textbf{5.68} &      95.91     &     0.0403      &  \textbf{21.42} &      97.68      &      88.37     &  24.77  &       63.31    \\ \hline


\end{tabular}
}

\caption{
TESSP evaluation on SUPERB benchmark. ParaL denotes paralinguistic aspect of speech.  \label{table:superb}
}
\end{table*}

\begin{table*}[h]
\begin{center}
\begin{small}
\resizebox{1.0\textwidth}{!}{
\begin{tabular}{l|ccccc|cc|cc|cc}
\hline
\multirow{2}{*}{Model} & \multicolumn{5}{c|}{Pre-training} & \multicolumn{2}{c|}{Fine-tuning \& decoding } & \multicolumn{2}{c|}{WER$\%$($\downarrow$) w/o LM} & \multicolumn{2}{c}{WER$\%$($\downarrow$) w/ LM} \\
 & $L_{\text{CTC}}$ & $L_{\text{MLM}}$ &\#Paired data & emb() & Char layer & Char layer & Char head & test-clean & test-other & test-clean & test-other \\ \hline
\hline
WavLM & -          &    -       & -      & -    & - & - &  - & 5.7  & 12.0 & 3.4 & 7.7 \\
TESSP & \Checkmark & \Checkmark & 100 hr & swap & \Checkmark &\XSolidBrush & \XSolidBrush & 5.10 & 10.88 & 3.05 & 7.15 \\ 

\hline
\multicolumn{11}{l}{\textbf{\textit{Label efficiency}}} \\ \hline
TESSP & \XSolidBrush & \Checkmark & 100 hr & swap & \Checkmark & \XSolidBrush & \XSolidBrush & 5.34 & 11.50 & 3.24 & 7.67 \\
TESSP &\Checkmark & \XSolidBrush & 100 hr & swap & \Checkmark &\XSolidBrush  & \XSolidBrush & 5.29 & 11.38 & 3.22 & 7.58\\ 

\hline
\multicolumn{11}{l}{\textbf{\textit{Paired efficiency}}} \\ \hline
TESSP & \Checkmark & \Checkmark & 0 hr & swap & \Checkmark & \XSolidBrush & \XSolidBrush & 5.7  & 12.2 & 3.44 & 7.90  \\ 
TESSP & \Checkmark & \Checkmark & 1 hr & swap & \Checkmark & \XSolidBrush & \XSolidBrush & 5.10 & 11.17 & 3.24 & 7.57 \\

\hline
\multicolumn{11}{l}{\textbf{\textit{Alignment function efficiency}}} \\ \hline
TESSP & \Checkmark & \Checkmark & 100 hr & CE loss & \Checkmark & \XSolidBrush & \XSolidBrush & 5.10 & 11.62 & 3.36 & 7.71\\ 
TESSP & \Checkmark & \Checkmark & 100 hr & cross attention & \Checkmark & \XSolidBrush & \XSolidBrush & 5.61 & 11.70 & 3.2 & 7.61\\ 
\hline
\multicolumn{11}{l}{\textbf{\textit{Character encoding layer efficiency (Zero shot)}}} \\ \hline
TESSP & \Checkmark & \Checkmark & 100 hr & swap & \Checkmark & \Checkmark & \Checkmark & 73.75 & 79.39 & 71.30 & 78.69  \\ 
TESSP & \Checkmark & \Checkmark & 100 hr & swap & \XSolidBrush & \XSolidBrush & \Checkmark & 103.16 & 106.08 & 103.16 & 106.08 \\ 
\hline
\multicolumn{11}{l}{\textbf{\textit{Character head layer efficiency }}} \\\hline
TESSP & \Checkmark & \Checkmark & 100 hr & swap & \Checkmark & \Checkmark & \Checkmark & 3.92 & 9.15 & 2.80 & 6.58 \\ 
TESSP & \Checkmark & \Checkmark & 100 hr & swap & \Checkmark & \Checkmark & \XSolidBrush & 4.72 & 10.31 & 3.06 & 7.02 \\ 
TESSP & \Checkmark & \Checkmark & 100 hr & swap & \XSolidBrush & \XSolidBrush & \Checkmark & 5.28 & 11.27 & 3.31 & 7.67 \\ 
TESSP & \Checkmark & \Checkmark & 100 hr & swap & \XSolidBrush & \XSolidBrush & \XSolidBrush & 4.93 & 10.97 & 3.13 & 7.42  \\ \hline
\end{tabular}
}
\caption{Ablation study on Librispeech dataset. \#Paired data denotes the dataset size for paired data while emb() denotes the embedding process method before shared encoder during training. Char layer denotes using character encoding layer in pre-training, fine-tuning or decoding. Char head denotes using pre-trained character linear embedding in fine-tuning or decoding. The language model used here is 4-gram language model. }
\label{ablabtion study}
\end{small}
\end{center}
\vspace{-1.0em}
\end{table*}

\subsection{Evaluation on Librispeech}
We compare our method with several competitive self-supervised approaches in the literature, including DeCoAR 2.0~\cite{ling2020decoar}, DiscreteBERT~\cite{baevski2019effectiveness}, wav2vec 2.0~\cite{baevski2020wav2vec}, HuBERT~\cite{hsu2021hubert} and WavLM~\cite{chen2022wavlm}.  As HuBERT only reports the WER with LM in their paper, we fine-tune their released model on supervised sets and report the without LM results. Table~\ref{main result} presents the results of our TESSP model, in which the model architecture follows the WavLM \textsc{Base} setting. On the test-clean set, our method reaches relative 10.5\% WER reductions against the WavLM baseline. The gain is even larger on the noisy test-other set, where the relative WER reduction is 11.5\%. With the 4-gram LM fusion, TESSP outperforms WavLM by 10.29\% and 11.7\% relatively. This indicates that our pre-training method can learn a better speech representation from additional text data.

\subsection{Evaluation on SUPERB}
Recent work~\cite{chen2022wavlm} shows that speech pre-trained models can solve full stack speech processing tasks. As shown in Table~\ref{table:superb}, we evaluate the proposed method on the SUPERB benchmark~\cite{yang2021superb} which is designed to provide a standard and comprehensive test board for pre-trained models on various speech tasks. It contains thirteen tasks, including Speaker Identification (SID), Automatic Speaker Verification (ASV), Speaker Diarization (SD), Phoneme Recognition (PR), Automatic Speech Recognition (ASR), Keyword Spotting (KS), Query by Example Spoken Term Detection (QbE), Speech Translation (ST), Intent Classification (IC), Slot Filling(SF), Emotion Recognition(ER). These tasks can be grouped into four aspects of speech: content, speaker, semantics and paralinguistics.

Table~\ref{table:superb} shows the performance of TESSP on SUPERB. Our model is compared with wav2vec~\cite{baevski2020wav2vec}, vq-wav2vec~\cite{baevski2019vq}, wav2vec 2.0, HuBERT and WavLM. TESSP shows better performance than HuBERT BASE on PR, ASR and ST tasks, while performance degradation for other tasks. Due to the influence of unsupervised text data, our model is more friendly to ASR-related tasks which require strong contextual information. For KS and QbE mainly find the inherent pattern in the speech signal, the two tasks can not get improved from unsupervised text data.

\section{Analysis \& Discussion}

\subsection{Ablation study}

\subsubsection{Text Label Efficiency}
Table~\ref{ablabtion study} shows the ablation study on the loss efficiency. Our model without character-level CTC loss performs 4.7\% and 5.7\% descend on test-clean and test-other sets. We also find that without phoneme-level MLM loss, even adding character-level CTC loss at the 50000 steps, it will still be difficult for CTC loss to converge. Although with the decline of the learning rate, CTC loss converges incompletely, the model performs 3.7\% and 4.6\% descend on test-clean and test-other sets, respectively.

\subsubsection{Paired Data Efficiency\label{no paired model}}

During speech and text joint pre-training, we use aligned paired speech and phoneme data to introduce alignment information. To verify whether the extra alignment information is necessary for the model, we train the TESSP without supervised training and the result shows no improvement over the baseline WavLM model. We further leverage 1-hour labeled data which is randomly selected from 100 hours of labeled data in paired data pre-training. The result shows nearly performance which shows that TESSP only needs a small set of supervised data to help the model learn the alignment.

\subsubsection{Alignment Function Efficiency}

We mentioned above that the addition of paired data can greatly improve the performance of the model. But should we leverage the paired data explicitly or implicitly? We compared two main methods:1. swap at the corresponding position between speech and phoneme sequence as mentioned above, 2. introduce the cross entropy (CE) loss between the hidden state of speech and phoneme sequence on unmasked position. Table~\ref{ablabtion study} shows that the CE loss performs the same as the swap method on the test-clean set while increasing 6.8\% WER on the test-other set relatively. The result shows that the swap method is more efficient than the CE loss in the noisy scenario.

\subsubsection{Character Encoding Layer \& Character Head Efficiency}
As mentioned in Section~\ref{model architecture}, we add an external Transformer layer, called character encoding layer, and a character-level linear embedding, called character head, into TESSP to learn the character-level representation. We further study the efficiency of the two modules. Cause we softly add the alignment information by swapping embedding during paired data pre-training, TESSP does not know how to transcribe the speech sequence to text sequence. So we first evaluate the character encoding layer efficiency without fine-tuning the model. For the model without character encoding layer, the output $H^t$ of shared encoder is directly used to calculate $L_{\text{MLM}}$ and $L_{\text{CTC}}$.
Results in Table~\ref{ablabtion study} show that TESSP has a weak ability to generate correct characters without fine-tuning. However, without the extra character encoding layer, the model loses this ability which shows that the character encoding layer helps the model learn the character information.

To study the character head efficiency, we fine-tune the model with or without the character head for both the TESSP and TESSP without character encoding layer model. As for TESSP, we find that the model fine-tuned with the pre-trained character head outperforms it fine-tuned without the pre-trained character 16.9\% and 11\% on test-clean and test-other relatively. However, the result of TESSP without character encoding layer is totally opposite. The result of without fine-tuning experiments has proved TESSP learns bad character representations without character encoding layer. So the pre-trained character head learning from a bad character representation increases the training difficulty. 

Another interesting finding is that the TESSP fine-tuned without character encoding layer and character head has good performance no matter pre-trained with or without character encoding layer. As mentioned above, TESSP pre-trained without character encoding layer learns bad character representations. Analogously, TESSP pre-trained with character encoding layer learns phoneme-level representation on 12-th layer. As a conclusion, the two models both do not learn the character-level representation for speech input which makes them perform similarly in ASR task. What exactly affects TESSP have good performance in ASR task? We will answer this question in Section~\ref{alignment analysis}.
\begin{figure}[t]
\centering
\includegraphics[width=1.0\columnwidth]{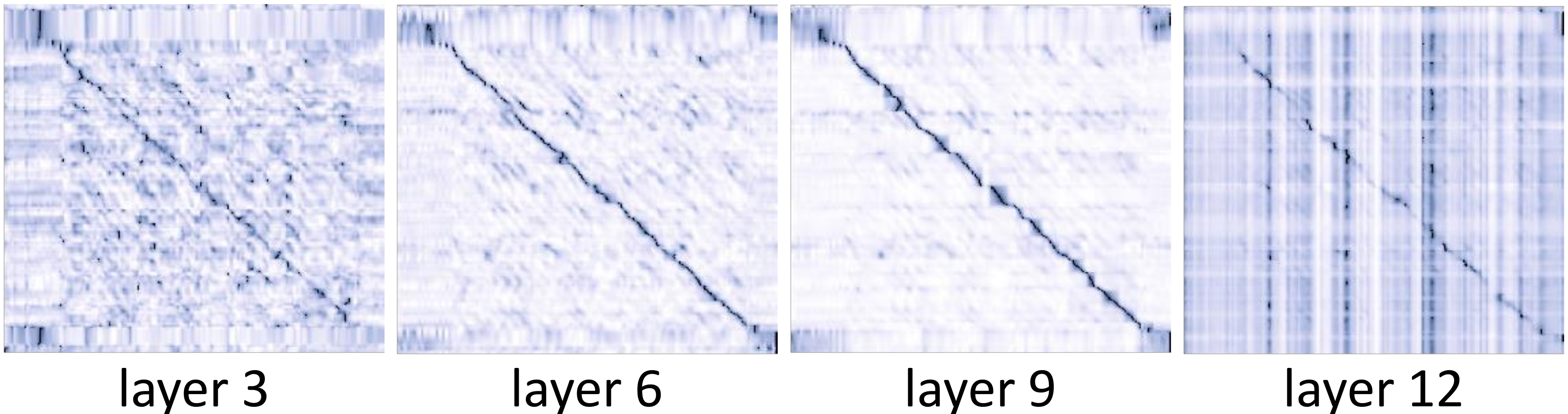} 
\caption{The normalized cosine similarity between speech and text data at different layers. \textit{layer 3} and \textit{layer 6} belong to the speech and text encoder respectively while \textit{layer 9} and \textit{layer 12} belong to the shared encoder. }
\label{fig3}
\vspace{-1.5em}
\end{figure}
\subsection{Analysis}
As mentioned above, TESSP learns a bad representation for the character. We further study the reason for better performance on the alignment between speech and text representation and the representation space between speech and text.

\subsubsection{Alignment between Speech and Text \label{alignment analysis}}
To verify that our model learns the alignment of speech and text data, we visualize the correlation between speech sequence and up-sampled phoneme sequence at different Transformer layers on LibriSpeech dev-clean subset. We compute the cosine similarity between the hidden states of aligned speech-phoneme pair data. Then we use bi-linear interpolation to resize $N$ different hot maps from $N$ utterances. We draw the final hot map by normalizing $N$ different hot maps with the max-min normalization method and average them. In our experiment, the $l \in {3,6,9,12}$-th layer of the Transformer encoder is set to visualize and $N=200$ utterances of speech-phoneme pairs were randomly selected.

\begin{figure}[t]
\centering
\includegraphics[width=1.0\columnwidth]{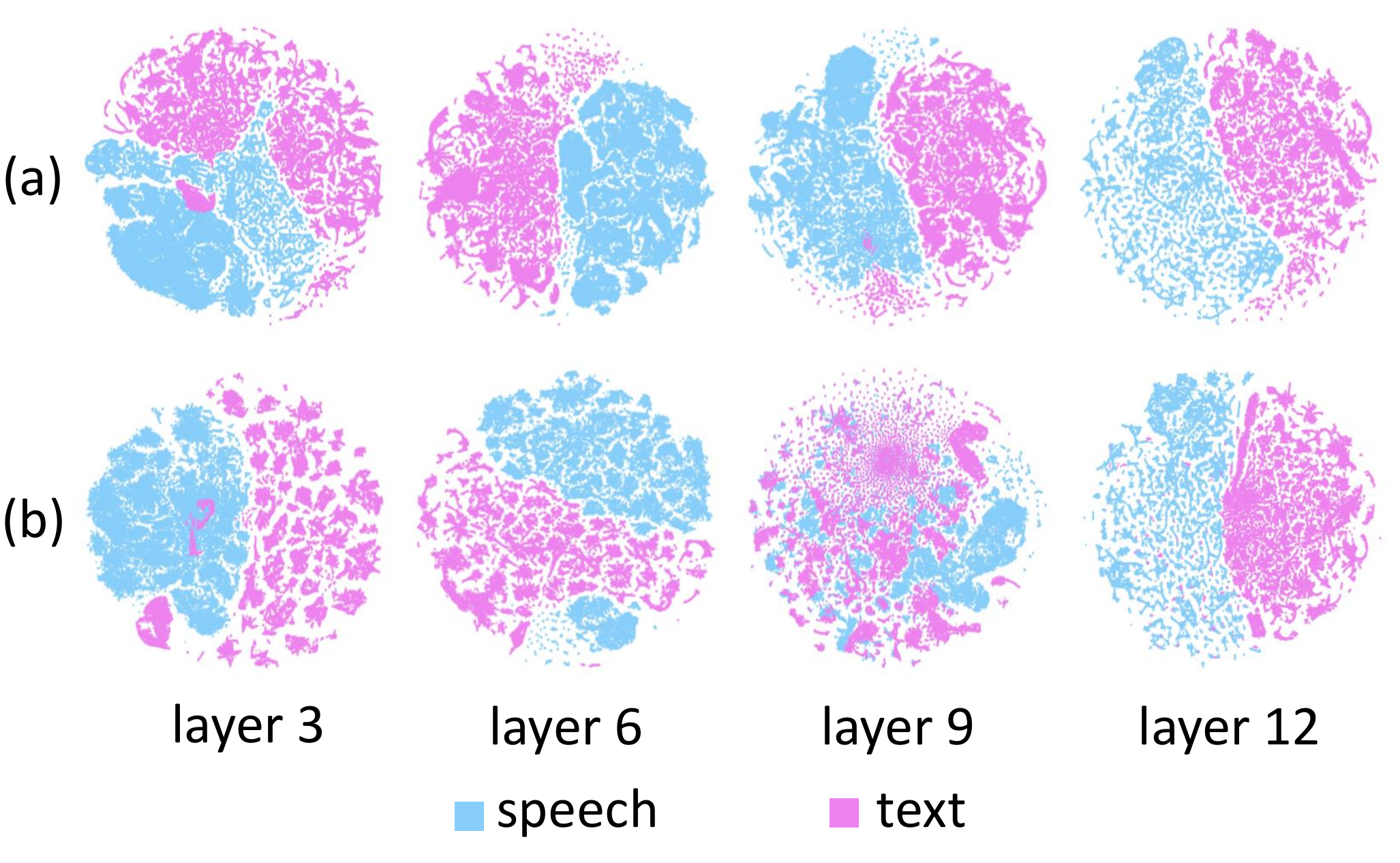} 
\vspace{-2.0em}
\caption{2-D illustration of the token-level representations of TESSP. Representations drawn in (a) are extracted from TESSP without paired data pre-training as described in Section~\ref{no paired model}, while (b) is for the TESSP model.}
\vspace{-1.0em}
\label{fig4}
\end{figure}
The result is shown in Figure~\ref{fig3}. From Figure~\ref{fig3}, we can see a diagonal line at each selected layer, which shows the alignment between speech and text data learned by the model. The diagonal line is gradually clear from layer 3 to layer 9 and becomes blurred at layer 12. We will give a detailed discussion about it in the next section.
\subsubsection{Speech and Text Representation Space}
Since the proposed TESSP aims to enhance the speech representation by text data, we suppose that it is able to learn a shared representation space of speech and text. 
To verify our assumption, we randomly sample 100 utterances from the representations extracted in the last section. The same method is used to extract the representations from the model described in Section~\ref{no paired model}. T-SNE~\cite{van2008visualizing} is performed to reduce the dimension to 2D.
Figure~\ref{fig4} (a) shows that the model can not learn the shared representation space for the two modalities without the paired data.
Figure~\ref{fig4} (b) shows that the representations of TESSP are gradually fused as the layer increases but divided into two regions at layer 12. This phenomenon explains why the diagonal line gets blurred at layer 12 in Figure~\ref{fig3}. An explanation is that the HuBERT target and phoneme target perform in different distributions. While TESSP tries to learn the shared representation of the two modalities through the alignment information from paired data, TESSP learns the respective knowledge of the two modalities in the high layers. However, TESSP still has good performance in downstream ASR tasks. The analysis shows that TESSP learns the alignment between speech and text data and the shared representation in the middle layer which helps TESSP get better performance in the down-streaming ASR task.

\section{Conclusion}
In this paper, we propose TESSP to jointly pre-training the speech data and text data, which outperforms WavLM by 10.29\% and 11.7\% on down streaming ASR task on Librispeech dataset. Results shows on SUPERB benchmark that the method is efficient to some content tasks. Analysis shows our method can learn reasonable alignment between speech and text data. Analysis experiments show the model learns the shared representation and alignment of speech and text which make TESSP efficiency. However, the model does not learn the character representation during pre-training which will be studied to solve in future work.

\bibliography{aaai23}

\begin{thebibliography}{40}
\providecommand{\natexlab}[1]{#1}

\bibitem[{Ao et~al.(2022)Ao, Wang, Zhou, Wang, Ren, Wu, Liu, Ko, Li, Zhang,
  Wei, Qian, Li, and Wei}]{ao2021speecht5}
Ao, J.; Wang, R.; Zhou, L.; Wang, C.; Ren, S.; Wu, Y.; Liu, S.; Ko, T.; Li, Q.;
  Zhang, Y.; Wei, Z.; Qian, Y.; Li, J.; and Wei, F. 2022.
\newblock SpeechT5: Unified-Modal Encoder-Decoder Pre-Training for Spoken
  Language Processing.
\newblock In \emph{Proc. of ACL}.

\bibitem[{Baevski, Auli, and Mohamed(2019)}]{baevski2019effectiveness}
Baevski, A.; Auli, M.; and Mohamed, A. 2019.
\newblock Effectiveness of self-supervised pre-training for speech recognition.

\bibitem[{Baevski et~al.(2021)Baevski, Hsu, Conneau, and
  Auli}]{baevski2021unsupervised}
Baevski, A.; Hsu, W.-N.; Conneau, A.; and Auli, M. 2021.
\newblock Unsupervised speech recognition.
\newblock \emph{Proc. of NeurIPS}.

\bibitem[{Baevski et~al.(2022)Baevski, Hsu, Xu, Babu, Gu, and
  Auli}]{baevski2022data2vec}
Baevski, A.; Hsu, W.-N.; Xu, Q.; Babu, A.; Gu, J.; and Auli, M. 2022.
\newblock Data2vec: A general framework for self-supervised learning in speech,
  vision and language.

\bibitem[{Baevski, Schneider, and Auli(2019)}]{baevski2019vq}
Baevski, A.; Schneider, S.; and Auli, M. 2019.
\newblock vq-wav2vec: Self-Supervised Learning of Discrete Speech
  Representations.
\newblock In \emph{Proc. of ICLR}.

\bibitem[{Baevski et~al.(2020)Baevski, Zhou, Mohamed, and
  Auli}]{baevski2020wav2vec}
Baevski, A.; Zhou, Y.; Mohamed, A.; and Auli, M. 2020.
\newblock wav2vec 2.0: A framework for self-supervised learning of speech
  representations.
\newblock \emph{Proc. of NeurIPS}.

\bibitem[{Bapna et~al.(2022)Bapna, Cherry, Zhang, Jia, Johnson, Cheng, Khanuja,
  Riesa, and Conneau}]{bapna2022mslam}
Bapna, A.; Cherry, C.; Zhang, Y.; Jia, Y.; Johnson, M.; Cheng, Y.; Khanuja, S.;
  Riesa, J.; and Conneau, A. 2022.
\newblock mSLAM: Massively multilingual joint pre-training for speech and text.

\bibitem[{Bapna et~al.(2021)Bapna, Chung, Wu, Gulati, Jia, Clark, Johnson,
  Riesa, Conneau, and Zhang}]{bapna2021slam}
Bapna, A.; Chung, Y.-a.; Wu, N.; Gulati, A.; Jia, Y.; Clark, J.~H.; Johnson,
  M.; Riesa, J.; Conneau, A.; and Zhang, Y. 2021.
\newblock SLAM: A unified encoder for speech and language modeling via
  speech-text joint pre-training.

\bibitem[{Bengio et~al.(2003)Bengio, Ducharme, Vincent, and
  Jauvin}]{bengio2003neural}
Bengio, Y.; Ducharme, R.; Vincent, P.; and Jauvin, C. 2003.
\newblock A Neural Probabilistic Language Model.
\newblock \emph{Journal of Machine Learning Research}.

\bibitem[{Chen et~al.(2022{\natexlab{a}})Chen, Wang, Chen, Wu, Liu, Chen, Li,
  Kanda, Yoshioka, Xiao et~al.}]{chen2022wavlm}
Chen, S.; Wang, C.; Chen, Z.; Wu, Y.; Liu, S.; Chen, Z.; Li, J.; Kanda, N.;
  Yoshioka, T.; Xiao, X.; et~al. 2022{\natexlab{a}}.
\newblock Wavlm: Large-scale self-supervised pre-training for full stack speech
  processing.
\newblock \emph{IEEE Journal of Selected Topics in Signal Processing}.

\bibitem[{Chen et~al.(2022{\natexlab{b}})Chen, Zhang, Rosenberg, Ramabhadran,
  Moreno, Bapna, and Zen}]{chen2022maestro}
Chen, Z.; Zhang, Y.; Rosenberg, A.; Ramabhadran, B.; Moreno, P.; Bapna, A.; and
  Zen, H. 2022{\natexlab{b}}.
\newblock MAESTRO: Matched Speech Text Representations through Modality
  Matching.

\bibitem[{Chi et~al.(2022)Chi, Huang, Dong, Ma, Zheng, Singhal, Bajaj, Song,
  Mao, Huang et~al.}]{chi2022xlm}
Chi, Z.; Huang, S.; Dong, L.; Ma, S.; Zheng, B.; Singhal, S.; Bajaj, P.; Song,
  X.; Mao, X.-L.; Huang, H.-Y.; et~al. 2022.
\newblock XLM-E: Cross-lingual Language Model Pre-training via ELECTRA.
\newblock In \emph{Proc. of ACL}.

\bibitem[{Chung et~al.(2016)Chung, Wu, Shen, Lee, and Lee}]{chung2016audio}
Chung, Y.-A.; Wu, C.-C.; Shen, C.-H.; Lee, H.-Y.; and Lee, L.-S. 2016.
\newblock Audio Word2Vec: Unsupervised Learning of Audio Segment
  Representations Using Sequence-to-Sequence Autoencoder.
\newblock In \emph{Proc. of Interspeech}.

\bibitem[{Devlin et~al.(2019)Devlin, Chang, Lee, and
  Toutanova}]{devlin2018bert}
Devlin, J.; Chang, M.-W.; Lee, K.; and Toutanova, K. 2019.
\newblock {BERT}: Pre-training of Deep Bidirectional Transformers for Language
  Understanding.
\newblock In \emph{Proc. of NAACL}.

\bibitem[{Graves et~al.(2006)Graves, Fern{\'a}ndez, Gomez, and
  Schmidhuber}]{graves2006connectionist}
Graves, A.; Fern{\'a}ndez, S.; Gomez, F.; and Schmidhuber, J. 2006.
\newblock Connectionist temporal classification: labelling unsegmented sequence
  data with recurrent neural networks.
\newblock In \emph{Proc. of ICML}.

\bibitem[{Gutmann and Hyv{\"a}rinen(2012)}]{gutmann2012noise}
Gutmann, M.~U.; and Hyv{\"a}rinen, A. 2012.
\newblock Noise-Contrastive Estimation of Unnormalized Statistical Models, with
  Applications to Natural Image Statistics.
\newblock \emph{Journal of Machine Learning Research}.

\bibitem[{Hsu et~al.(2021)Hsu, Bolte, Tsai, Lakhotia, Salakhutdinov, and
  Mohamed}]{hsu2021hubert}
Hsu, W.-N.; Bolte, B.; Tsai, Y.-H.~H.; Lakhotia, K.; Salakhutdinov, R.; and
  Mohamed, A. 2021.
\newblock Hubert: Self-supervised speech representation learning by masked
  prediction of hidden units.
\newblock \emph{IEEE/ACM Transactions on Audio, Speech, and Language
  Processing}.

\bibitem[{Hu et~al.(2020)Hu, Ruder, Siddhant, Neubig, Firat, and
  Johnson}]{hu2020xtreme}
Hu, J.; Ruder, S.; Siddhant, A.; Neubig, G.; Firat, O.; and Johnson, M. 2020.
\newblock Xtreme: A massively multilingual multi-task benchmark for evaluating
  cross-lingual generalisation.
\newblock In \emph{Proc. of ICML}.

\bibitem[{Jang, Gu, and Poole(2017)}]{jang2017categorical}
Jang, E.; Gu, S.; and Poole, B. 2017.
\newblock Categorical Reparametrization with Gumble-Softmax.
\newblock In \emph{Proc. of ICLR}.

\bibitem[{Kim et~al.(2021)Kim, Kim, Lee, and Ha}]{kim2021st}
Kim, M.; Kim, G.; Lee, S.-W.; and Ha, J.-W. 2021.
\newblock St-Bert: Cross-Modal Language Model Pre-Training for End-to-End
  Spoken Language Understanding.
\newblock In \emph{Proc. of ICASSP}.

\bibitem[{Lewis et~al.(2020)Lewis, Liu, Goyal, Ghazvininejad, Mohamed, Levy,
  Stoyanov, and Zettlemoyer}]{lewis2020bart}
Lewis, M.; Liu, Y.; Goyal, N.; Ghazvininejad, M.; Mohamed, A.; Levy, O.;
  Stoyanov, V.; and Zettlemoyer, L. 2020.
\newblock BART: Denoising Sequence-to-Sequence Pre-training for Natural
  Language Generation, Translation, and Comprehension.
\newblock In \emph{Proc. of ACL}.

\bibitem[{Ling and Liu(2020)}]{ling2020decoar}
Ling, S.; and Liu, Y. 2020.
\newblock DeCoAR 2.0: Deep Contextualized Acoustic Representations with Vector
  Quantization.

\bibitem[{Liu et~al.(2022)Liu, Hsu, Auli, and Baevski}]{liu2022towards}
Liu, A.~H.; Hsu, W.-N.; Auli, M.; and Baevski, A. 2022.
\newblock Towards End-to-end Unsupervised Speech Recognition.

\bibitem[{Mikolov et~al.(2013)Mikolov, Sutskever, Chen, Corrado, and
  Dean}]{mikolov2013distributed}
Mikolov, T.; Sutskever, I.; Chen, K.; Corrado, G.~S.; and Dean, J. 2013.
\newblock Distributed representations of words and phrases and their
  compositionality.
\newblock \emph{Proc. of NeurIPS}.

\bibitem[{Mnih and Teh(2012)}]{mnih2012fast}
Mnih, A.; and Teh, Y.~W. 2012.
\newblock A fast and simple algorithm for training neural probabilistic
  language models.
\newblock In \emph{Proc. of ICML}.

\bibitem[{Oord, Li, and Vinyals(2018)}]{oord2018representation}
Oord, A. v.~d.; Li, Y.; and Vinyals, O. 2018.
\newblock Representation learning with contrastive predictive coding.

\bibitem[{Panayotov et~al.(2015)Panayotov, Chen, Povey, and
  Khudanpur}]{panayotov2015librispeech}
Panayotov, V.; Chen, G.; Povey, D.; and Khudanpur, S. 2015.
\newblock Librispeech: an asr corpus based on public domain audio books.
\newblock In \emph{Proc. of ICASSP}.

\bibitem[{Pratap et~al.(2019)Pratap, Hannun, Xu, Cai, Kahn, Synnaeve,
  Liptchinsky, and Collobert}]{pratap2019wav2letter++}
Pratap, V.; Hannun, A.; Xu, Q.; Cai, J.; Kahn, J.; Synnaeve, G.; Liptchinsky,
  V.; and Collobert, R. 2019.
\newblock Wav2letter++: A fast open-source speech recognition system.
\newblock In \emph{Proc. of ICASSP}.

\bibitem[{Qian et~al.(2021)Qian, Bianv, Shi, Kanda, Shen, Xiao, and
  Zeng}]{qian2021speech}
Qian, Y.; Bianv, X.; Shi, Y.; Kanda, N.; Shen, L.; Xiao, Z.; and Zeng, M. 2021.
\newblock Speech-language pre-training for end-to-end spoken language
  understanding.
\newblock In \emph{Proc. of ICASSP}.

\bibitem[{Radford et~al.(2018)Radford, Narasimhan, Salimans, Sutskever
  et~al.}]{radford2018improving}
Radford, A.; Narasimhan, K.; Salimans, T.; Sutskever, I.; et~al. 2018.
\newblock Improving language understanding by generative pre-training.

\bibitem[{Raffel et~al.(2020)Raffel, Shazeer, Roberts, Lee, Narang, Matena,
  Zhou, Li, Liu et~al.}]{raffel2020exploring}
Raffel, C.; Shazeer, N.; Roberts, A.; Lee, K.; Narang, S.; Matena, M.; Zhou,
  Y.; Li, W.; Liu, P.~J.; et~al. 2020.
\newblock Exploring the limits of transfer learning with a unified text-to-text
  transformer.
\newblock \emph{Journal of Machine Learning Research}.

\bibitem[{Sarzynska-Wawer et~al.(2021)Sarzynska-Wawer, Wawer, Pawlak,
  Szymanowska, Stefaniak, Jarkiewicz, and Okruszek}]{sarzynska2021detecting}
Sarzynska-Wawer, J.; Wawer, A.; Pawlak, A.; Szymanowska, J.; Stefaniak, I.;
  Jarkiewicz, M.; and Okruszek, L. 2021.
\newblock Detecting formal thought disorder by deep contextualized word
  representations.
\newblock \emph{Psychiatry Research}.

\bibitem[{Schneider et~al.(2019)Schneider, Baevski, Collobert, and
  Auli}]{schneider2019wav2vec}
Schneider, S.; Baevski, A.; Collobert, R.; and Auli, M. 2019.
\newblock wav2vec: Unsupervised Pre-Training for Speech Recognition.
\newblock In \emph{Proc. of Interspeech}.

\bibitem[{Taylor(1953)}]{taylor1953cloze}
Taylor, W.~L. 1953.
\newblock “Cloze procedure”: A new tool for measuring readability.
\newblock \emph{Journalism Quarterly}.

\bibitem[{Van Den~Oord, Vinyals et~al.(2017)}]{van2017neural}
Van Den~Oord, A.; Vinyals, O.; et~al. 2017.
\newblock Neural discrete representation learning.
\newblock \emph{Proc. of NeurIPS}.

\bibitem[{Van~der Maaten and Hinton(2008)}]{van2008visualizing}
Van~der Maaten, L.; and Hinton, G. 2008.
\newblock Visualizing data using t-SNE.
\newblock \emph{Journal of Machine Learning Research}.

\bibitem[{Wang et~al.(2019)Wang, Pruksachatkun, Nangia, Singh, Michael, Hill,
  Levy, and Bowman}]{wang2019superglue}
Wang, A.; Pruksachatkun, Y.; Nangia, N.; Singh, A.; Michael, J.; Hill, F.;
  Levy, O.; and Bowman, S. 2019.
\newblock Superglue: A stickier benchmark for general-purpose language
  understanding systems.
\newblock \emph{Proc. of NeurIPS}.

\bibitem[{Wang et~al.(2022)Wang, Ren, Qian, Liu, Shi, Qian, and
  Zeng}]{wang2022optimizing}
Wang, W.; Ren, S.; Qian, Y.; Liu, S.; Shi, Y.; Qian, Y.; and Zeng, M. 2022.
\newblock Optimizing Alignment of Speech and Language Latent Spaces for
  End-to-End Speech Recognition and Understanding.
\newblock In \emph{Proc. of ICASSP}.

\bibitem[{Yang et~al.(2021)Yang, Chi, Chuang, Lai, Lakhotia, Lin, Liu, Shi,
  Chang, Lin et~al.}]{yang2021superb}
Yang, S.-w.; Chi, P.-H.; Chuang, Y.-S.; Lai, C.-I.~J.; Lakhotia, K.; Lin,
  Y.~Y.; Liu, A.~T.; Shi, J.; Chang, X.; Lin, G.-T.; et~al. 2021.
\newblock Superb: Speech processing universal performance benchmark.

\bibitem[{Zheng et~al.(2021)Zheng, Xiao, Gong, Zhou, Liang, and
  Lin}]{zheng2021wav}
Zheng, G.; Xiao, Y.; Gong, K.; Zhou, P.; Liang, X.; and Lin, L. 2021.
\newblock Wav-BERT: Cooperative Acoustic and Linguistic Representation Learning
  for Low-Resource Speech Recognition.
\newblock In \emph{Proc. of EMNLP}.

\end{thebibliography}

\end{document}